\documentclass[preprint,12pt]{elsarticle}

\usepackage{amsmath,euscript,amssymb,epsfig,amsfonts,latexsym}
\usepackage{graphicx,color}

\newcommand{\be}[1]{\begin{equation}\label{#1}}
\newcommand{\ee}{\end{equation}}
\newcommand{\ba}[1]{\begin{eqnarray}\label{#1}}
\newcommand{\ea}{\end{eqnarray}}
\newcommand{\rf}[1]{(\ref{#1})}
\newcommand{\nn}{\nonumber}

\begin{document}

\begin{frontmatter}

\title{Duel of cosmological screening lengths}

\author{Ezgi Canay}\ead{ezgicanay@itu.edu.tr}

\address{Department of Physics, Istanbul Technical University, 34469 Maslak, Istanbul, Turkey}

\author{Maxim Eingorn}\ead{maxim.eingorn@gmail.com}

\address{Department of Mathematics and Physics, North Carolina Central University,\\ 1801 Fayetteville St., Durham, North Carolina 27707, U.S.A.}

\begin{abstract}
Two distinct perturbative approaches have been recently formulated within General Relativity, arguing for the screening of gravity in the $\Lambda$CDM Universe. In this paper we compare them and show that the offered screening concepts, each characterized by its own interaction range, can peacefully coexist. Accordingly, we advance a united scheme, determining the gravitational potential at all scales, including regions of nonlinear density contrasts, by means of a simple Helmholtz equation with the effective cosmological screening length. In addition, we claim that cosmic structures may not grow at distances above this Yukawa range and confront its current value with dimensions of the largest known objects in the Universe.
\end{abstract}


\end{frontmatter}

\

\section{Introduction}
\label{Sec1}

In the weak field limit, Einstein's General Relativity predicts Newtonian gravitational interaction between nonrelativistic massive bodies in the perturbed Minkowski spacetime \cite{Landau}, as well as in the perturbed Friedmann-Lema\^{\i}tre-Robertson-Walker spacetime at sub-horizon cosmological scales \cite{Peebles}. As the science rapidly progresses, higher and higher accuracy of observations is achieved. For instance, such a future space mission as Euclid \cite{Euclid1,Euclid2,Euclid3} is designed to probe the Universe expansion with unprecedented precision and impose new restrictions on dark energy, dark matter, and various cosmological parameters. In this connection, the legitimate demand for an advanced theory soars up too. In particular, it is quite natural to expect that Newtonian gravity is modified at large distances and ultimately reconciled with the standard relativistic perturbation theory. By dint of our narration, we aim at sparking interest in two distinct approaches \cite{Eingorn,Hahn} relying on General Relativity, which argue that gravity is actually screened, ceasing to be long-range far enough from its every single source. Hereinafter this Yukawa-type screening is sometimes called ``cosmological'' or ``cosmic'', since it originates from the presence of the cosmological background.

Both cosmic screening approaches have been formulated in the framework of the standard $\Lambda$CDM ($\Lambda$ cold dark matter) model, which is consistent with the observational data \cite{Planck}, though it is noteworthy that currently there is tension between the direct local measurement of the Hubble constant 
and the value of this very constant following from the Planck data on cosmic microwave background temperature and polarization (see, e.g., \cite{Hubten1,Hubten2,Hubten3}). Furthermore, both original papers \cite{Eingorn,Hahn} focus on the matter- and $\Lambda$-dominated stages of the Universe evolution, so radiation and relativistic neutrinos are disregarded (though the results of \cite{Eingorn} have been subsequently generalized to the case of additional perfect fluids with linear and nonlinear equations of state \cite{BrilEinPDU,Kiefer1,Kiefer2}, as well as to the cases of nonzero spatial curvature \cite{Emrah}, $f(R)$ gravity \cite{Shulga} and the phantom braneworld model \cite{Bhattacharya}).

From the mathematical point of view, the cosmological Yukawa screening inevitably comes into play when the Einstein equation for the gravitational potential (scalar perturbation) is reduced to the Helmholtz equation, which replaces its popular rival of Poisson type. Meanwhile, the underlying physical reasons and resulting screening ranges are different in \cite{Eingorn} and \cite{Hahn}. The scheme of \cite{Eingorn} (see additionally \cite{BrilEin,Duygu}) is rooted in the so-called discrete cosmology studying how discrete gravitating masses interact in the expanding Universe. The finite interaction range arises in \cite{Eingorn} and subsequent papers owing to the interpretation of the mass density fluctuation as the scalar perturbation source. Meanwhile, the scheme of \cite{Hahn} leans on the knowledge of the linear perturbation theory, and the screening scale arises owing to the weak temporal dependence of the gravitational potential.

The time has come to ask a couple of natural informal questions. Which screening length is better? And, more importantly, can two screening concepts coexist? Our current paper is devoted to these burning issues. Putting the cart before the horse, below we will ultimately answer, respectively: neither, yes!

The paper is organized as follows. In Sections~\ref{Sec2} and \ref{Sec3} we review the methods of discrete cosmology and linear perturbation theory, revealing the corresponding screening ranges. Moreover, we argue for the claim that the velocity-dependent source of the scalar perturbation should not be omitted, in spite of powerful incentives, rooted in deceptive numerical estimates. Then, in Section~\ref{Sec4}, we unite the schemes of \cite{Eingorn} and \cite{Hahn}, revealing the effective screening length, and study its properties. A brief summary of our main results is given in concluding Section~\ref{Sec5}.

\

\section{Screening length in discrete cosmology}
\label{Sec2}

According to \cite{Eingorn}, the inhomogeneous $\Lambda$CDM Universe can be quite accurately described at all scales (excluding regions of strong gravitational fields, where the weak field limit comes to grief) by the metric
\be{1} ds^2=a^2\left[\left(1 + 2\Phi\right)d\eta^2 + 2B_{\alpha}dx^{\alpha}d\eta - \left(1- 2\Phi\right)\delta_{\alpha\beta}dx^{\alpha}dx^{\beta}\right],\quad \alpha,\beta=1,2,3\, .\ee
Here $a(\eta)$ is the scale factor depending on the conformal time $\eta$ and satisfying the background Friedmann equations:
\be{2} \frac{3{\mathcal H}^2}{a^2}=\kappa\overline{\varepsilon} + \Lambda\, ,\quad \frac{2{\mathcal H}'+{\mathcal H}^2}{a^2}=\Lambda\, ,\ee
where ${\mathcal H}(\eta)\equiv (da/d\eta)/a\equiv a'/a$ (hereinafter the prime indicates the $\eta$-derivative), $\kappa\equiv 8\pi G_N/c^4$ ($G_N$ is the Newtonian gravitational constant while $c$ denotes the speed of light), $\overline{\varepsilon}(\eta)$ is the average energy density of nonrelativistic pressureless matter, and $\Lambda$ represents the cosmological constant. Returning to Eq.~\rf{1}, the metric corrections $\Phi(\eta,{\bf r})$ and $B_{\alpha}(\eta,{\bf r})$, depending on $\eta$ and the comoving coordinates $x^{\alpha}$ (here ${\bf r}\equiv\left(x^1,x^2,x^3\right)\equiv(x,y,z)$ stands for the radius-vector), are nothing else but the first-order scalar and vector perturbations, respectively. Tensor perturbations (cosmological gravitational waves) are not analyzed in \cite{Eingorn}, and we continue totally ignoring them in the current paper as well. One should mention that it is common practice to regard tensor modes as second-order perturbations (see, e.g., \cite{BrilEin,Baumann}).

If the inhomogeneous gravitational field is produced by a system of point-like particles (with comoving radius-vectors ${\bf r}_n(\eta)$, nonrelativistic peculiar velocities $\tilde v^{\alpha}_n(\eta)\equiv dx^{\alpha}_n/d\eta$, and masses $m_n$), then the analytical expressions for $\Phi$ and ${\bf B}\equiv(B_1,B_2,B_3)$ can be found from the corresponding Einstein equations (see Eqs.~(2.27) and (2.26) in \cite{Eingorn}, respectively):
\be{3} \triangle\Phi-\frac{3\kappa\overline\rho c^2}{2a}\Phi = \frac{\kappa c^2}{2a}\delta\rho-\frac{3\kappa c^2{\mathcal H}}{2a}\Xi
\, ,\ee
\be{4} \triangle{\bf B}-\frac{2\kappa\overline\rho c^2}{a}{\bf B}=-\frac{2\kappa c^2}{a}\left(\sum\limits_n\rho_n\tilde{\bf v}_n-\nabla\Xi\right)\, .\ee
Here $\triangle\equiv\delta^{\alpha\beta}\partial_{\alpha}\partial_{\beta}$, $\partial_{\alpha}\equiv\partial/\partial x^{\alpha}$, is the Laplace operator, and $\delta{\rho}(\eta,{\bf r})\equiv\rho-\overline{\rho}$ denotes the fluctuation of the mass density in comoving coordinates $\rho(\eta,{\bf r})$ around its constant average value $\overline{\rho}$ (which is naturally connected with $\overline{\varepsilon}$ via the equality $\overline\varepsilon=\overline\rho c^2/a^3\propto a^{-3}$). The considered system of point-like particles is characterized by the following expression for the mass density:
\be{5} \rho=\sum\limits_nm_n\delta({\bf r}-{\bf r}_n)=\sum\limits_n\rho_n\, ,\quad \rho_n\equiv m_n\delta({\bf r}-{\bf r}_n)\, .\ee

Returning to Eqs.~\rf{3} and \rf{4}, $\nabla{\Xi}$ is the longitudinal (curl-free) component of the vector $\sum_n\rho_n\tilde{\bf v}_n$, where $\tilde{\bf v}_n\equiv(\tilde{v}_n^1,\tilde{v}_n^2,\tilde{v}_n^3)$, so the function $\Xi(\eta,{\bf r})$ represents the solution of the Poisson equation $\triangle\Xi=\nabla\sum_n\rho_n\tilde {\bf v}_n$:
\be{6} \Xi=\frac{1}{4\pi}\sum\limits_nm_n\frac{({\bf r}-{\bf r}_n)\tilde {\bf v}_n}{|{\bf r}-{\bf r}_n|^3}\, .\ee

The exact solutions of the Helmholtz equations \rf{3}, \rf{4} have the following form (see Eqs.~(2.40) and (2.36) in \cite{Eingorn}, respectively):
\ba{7} \Phi&=&\frac{1}{3}-\frac{\kappa c^2}{8\pi a}\sum_{n}\frac{m_n}{|\mathbf{r}-\mathbf{r}_n|}\exp(-q_n)\nn\\
&+&\frac{3\kappa c^2\mathcal{H}}{8\pi a}\sum_{n}\frac{m_n[\tilde{\mathbf{v}}_n(\mathbf{r}-\mathbf{r}_n)]}{|\mathbf{r}-\mathbf{r}_n|}\,\frac{1-(1+q_n)\exp(-q_n)}{q^2_n}\, , \ea
\ba{8} {\bf B}&=& \frac{\kappa c^2}{8\pi a}\sum\limits_{n}\left[\frac{m_n\tilde{\bf v}_n}{|{\bf r}-{\bf r}_n|}\,
\frac{\left(3+2\sqrt{3}q_n+4q_n^2\right)\exp\left(-2q_n/\sqrt{3}\right)-3}{q_n^2}\right.\nn\\
&+&\left.\frac{m_n[\tilde{\bf v}_n({\bf r}-{\bf r}_n)]}{|{\bf r}-{\bf r}_n|^3}({\bf r}-{\bf r}_n)\, \frac{9-\left(9+6\sqrt{3}q_n+4q_n^2\right)\exp\left(-2q_n/\sqrt{3}\right)}{q_n^2}\right],\quad\,\ \ea
where $q_n(\eta,{\bf r})$ is the absolute value of the spatial vector ${\bf q}_n(\eta,{\bf r})$ defined as
\be{9} {\bf q}_n\equiv \sqrt{\frac{3\kappa\overline\rho c^2}{2a}}({\bf r}-{\bf r}_n)=\frac{a({\bf r}-{\bf r}_n)}{\lambda}\, ,\quad \lambda\equiv\sqrt{\frac{2a^3}{3\kappa\overline\rho c^2}}\, . \ee
Here $\lambda(\eta)\propto a^{3/2}$ is nothing else but the screening length defining the range of gravitational interaction in discrete cosmology. In contrast to \cite{Signore,EBV}, this finite range is not introduced by hand. Instead, it originates from the dependence of energy-momentum fluctuations on the metric corrections \cite{BrilEin}.

The analytical expressions \rf{7} and \rf{8} are valid for all spatial scales and perfectly conform with the Minkowski background limit and (sub-horizon) Newtonian cosmological approximation, as clearly demonstrated in \cite{Eingorn}. In particular, those regions, where $\rho\gg\overline\rho$ while the gravitational field remains weak, are wholly covered, similarly to, e.g., \cite{Baumann} (see \cite{Eingorn,BrilEin} for the detailed comparison with other perturbative approaches).

Being armed with Eqs.~\rf{7}, \rf{8}, let us write down the equation of motion of an arbitrary particle of the system under the influence of all other particles (see Eq.~(3.6) in \cite{Eingorn}):
\be{10} \left(a\tilde{\bf v}_k\right)'\ =\ -a\left(\nabla\Phi|_{{\bf r}={\bf r}_k}+\mathcal{H}{\bf B}|_{{\bf r}={\bf r}_k}\right)=\sum_{n\neq k} \mathbf{f}_n(\eta,\mathbf{r}_k)\, ,\ee
where $\mathbf{f}_n(\eta,\mathbf{r})$ denotes the force per unit mass, induced by the n-th particle. This spatial vector is defined as
\ba{11}
&{}&\mathbf{f}_n= -\frac{\kappa c^2}{8\pi}\left[\frac{m_n(\mathbf{r}-\mathbf{r}_n)}{|\mathbf{r}-\mathbf{r}_n|^3}\,(1+q_n)\exp(-q_n)\right. \, \nn\\
&+&\mathcal{H}\frac{m_n[\tilde{\mathbf{v}}_n(\mathbf{r}-\mathbf{r}_n)]}{|\mathbf{r}-\mathbf{r}_n|^3}(\mathbf{r}-\mathbf{r}_n)\nn\\
&\times&\left.\frac{9\left(1+q_n+q_n^2/3\right)\exp(-q_n)-\left(9+6\sqrt{3}q_n+4q_n^2\right)\exp\left(-2q_n/\sqrt{3}\right)}{q_n^2}\right.\, \nn\\
&+&\left.\mathcal{H}\frac{m_n\tilde{\mathbf{v}}_n}{|\mathbf{r}-\mathbf{r}_n|}\,\frac{\left(3+2\sqrt{3}q_n+4q_n^2\right)\exp\left(-2q_n/\sqrt{3}\right)-3(1+q_n)\exp(-q_n)}{q_n^2}
\right]\, .\nn\\
&{}&\ea

Obviously, there are terms without exponential functions in both Eqs.~\rf{7}, \rf{8}. Nevertheless, such terms do not survive in Eq.~\rf{11} in view of their irretrievable mutual cancellation. Thus, we have explicitly confirmed that the induced force decreases exponentially with distance from the corresponding particle \cite{Eingorn}.

Let us take a closer look at the formula \rf{11} and ask a very important question: are the velocity-dependent contributions negligible as compared with the velocity-independent one? Estimating the ratio ``$|$\,sum of two terms with $\tilde{\mathbf{v}}_n$\,$|$ / $|$\,lonely term without $\tilde{\mathbf{v}}_n$\,$|$'' for various $q_n$, one can show that the maximum value of this ratio is of the order of $3\mathcal{H}\lambda\tilde{v}_n/a=3H\lambda av_n/c^2$ (here $\tilde v_n\equiv|\tilde{\mathbf{v}}_n|$, $H=c\mathcal{H}/a$ stands for the Hubble parameter, $a{\mathbf{v}}_n=c\tilde{\mathbf{v}}_n$ is the physical peculiar velocity, $av_n\equiv|a{\mathbf{v}}_n|$). Using the current values of the Hubble parameter, screening length and typical peculiar velocity, $H_0\approx 70\,\mathrm{km}\,\mathrm{s}^{-1}\,\mathrm{Mpc}^{-1}$, $\lambda_0\approx3.7\,\mathrm{Gpc}$ and $(av_n)_0\sim 250\div500\,\mathrm{km}\,\mathrm{s}^{-1}$, respectively, we find that today $3H\lambda av_n/c^2\sim2\div4\times10^{-3}$. This means that the velocity-dependent part of ${\bf f}_n$ \rf{11} is much less than its velocity-free part \cite{Eingorn2017}.

Let us also ask the similar question with respect to the formula \rf{7}, momentarily ignoring the constant $1/3$: what is the ratio ``$|$\,single summand with $\tilde{\mathbf{v}}_n$ from the second line\,$|$ / $|$\,single summand without $\tilde{\mathbf{v}}_n$ from the first line\,$|$''? Since there is a term without exponential function in the second line of \rf{7} (containing $\tilde{\mathbf{v}}_n$), but no such term in the first line (being $\tilde{\mathbf{v}}_n$-free), it is quite logical to expect that at some distance from the particle the considered ratio becomes large. However, one can show that its maximum value remains small within the normally used cosmological simulation boxes: this value actually does not exceed $\sim1\div2\%$ at present for $q_n\leqslant3$ (i.e. for physical distances less than or equal to the homogeneity scale lower bound $\sim3\lambda_0\approx11\,\mathrm{Gpc}$ \cite{Eingorn,Li}).

Thus, it may seem that the velocity-dependent contributions are inessential for the cosmological simulation purposes, and one faces almost no risk of losing accuracy in rewriting the equation of motion \rf{10} in the much simpler form
\be{12} \left(a\tilde{\bf v}_k\right)'\ =\ -a\nabla\tilde\Phi|_{{\bf r}={\bf r}_k}\, ,\ee
where $\tilde\Phi(\eta,\mathbf{r})$ represents the $\tilde{\mathbf{v}}_n$-free part of $\Phi$ \rf{7}:
\be{13} \tilde\Phi=\frac{1}{3}-\frac{\kappa c^2}{8\pi a}\sum_{n}\frac{m_n}{|\mathbf{r}-\mathbf{r}_n|}\exp(-q_n)\, . \ee
Obviously, the function \rf{13} satisfies the Helmholtz equation, which, unlike Eq.~\rf{3}, has no source containing $\Xi$:
\be{14} \triangle\tilde\Phi-\frac{3\kappa\overline\rho c^2}{2a}\tilde\Phi = \frac{\kappa c^2}{2a}\delta\rho\, .\ee

Somewhat surprisingly, it turns out that the risk of losing accuracy is miscalculated, and the modification \rf{12} of the original equation of motion \rf{10} is a fatal mistake. Why? Because, as we will clearly demonstrate in the next section within the linear cosmological perturbation theory, this modification would lead to the wrong description of structure growth at sufficiently large scales. The reason is simple: the properties of the gravitational field generated by a solitary particle (point-like mass) should not be arrogated to accumulations of particles (distributed mass). In order to reveal the difference in the framework of discrete cosmology, we consider a ball of comoving radius $r_b$ and uniform mass density $\rho_b>\overline\rho$ as an illustrative example of the finite-size overdensity. According to \rf{14}, the volume element $d\mathcal{V}'$ of the ball (with radius-vector ${\mathbf r'}$) generates
\be{15} d\tilde\Phi_b=-\frac{\kappa c^2}{8\pi a}\frac{\left(\rho_b-\overline\rho\right)d\mathcal{V}'}{|\mathbf{r}-\mathbf{r'}|}\exp\left(-\frac{a|\mathbf{r}-\mathbf{r'}|}{\lambda}\right)\, .\ee
Integrating over the whole volume of the ball, for the outer region $r>r_b$ we find
\be{16}
\tilde\Phi_b=-\frac{\kappa c^2\lambda^3}{2 a^4}\frac{\rho_b-\overline{\rho}}{r}\left[\frac{ar_b}{\lambda}\cosh\left(\frac{ar_b}{\lambda}\right)-\sinh\left(\frac{ar_b}{\lambda}\right)\right]\exp\left(-\frac{ar}{\lambda}\right)\, ,
\ee
in complete agreement with the formula (3.7) from \cite{ball}.

Returning to Eq.~\rf{3} and its solution \rf{7}, in addition to $\tilde\Phi$ we introduce the function $\Phi_v(\eta,\mathbf{r})$, being the velocity-dependent part of $\Phi$:
\be{17} \Phi_v=\frac{3\kappa c^2\mathcal{H}}{8\pi a}\sum_{n}\frac{m_n[\tilde{\mathbf{v}}_n(\mathbf{r}-\mathbf{r}_n)]}{|\mathbf{r}-\mathbf{r}_n|}\,\frac{1-(1+q_n)\exp(-q_n)}{q^2_n}\, . \ee
This function satisfies the Helmholtz equation, which, unlike Eq.~\rf{3}, has no source containing $\delta\rho$:
\be{18} \triangle\Phi_v-\frac{3\kappa\overline\rho c^2}{2a}\Phi_v = -\frac{3\kappa c^2{\mathcal H}}{2a}\Xi\, .\ee

Let us imagine that the considered ball is moving as a whole with velocity $\tilde{\bf v}_b$, then the volume element $d\mathcal{V}'$ produces
\be{19}
d\Phi_{vb}=\frac{3\kappa c^2\mathcal{H}}{8\pi a}\frac{\rho_bd\mathcal{V}'}{|\mathbf{r}-\mathbf{r'}|}\left[\tilde{\mathbf{v}}_b(\mathbf{r}-\mathbf{r'})\right]\frac{1-(1+a|\mathbf{r}-\mathbf{r'}|/\lambda)\exp\left(-a|\mathbf{r}-\mathbf{r'}|/\lambda\right)}{a^2|\mathbf{r}-\mathbf{r'}|^2/\lambda^2}\, .
\ee
Assuming, for the sake of simplicity, that the vectors $\tilde{\bf v}_b$ and ${\bf r}$ are collinear, after exhausting integration we derive
\ba{20} &{}&\Phi_{vb}=-\frac{3\kappa c^2\mathcal{H}\lambda^5}{2a^6}\frac{\rho_b\tilde{v}_b}{r^2}\nn\\
&\times&\left\{-\frac{1}{3}\left(\frac{ar_b}{\lambda}\right)^3+\left(1+\frac{ar}{\lambda}\right)\left[\frac{ar_b}{\lambda}\cosh\left(\frac{ar_b}{\lambda}\right)-\sinh\left(\frac{ar_b}{\lambda}\right)\right]\exp\left(-\frac{ar}{\lambda}\right)\right\}\nn\\
&{}&\ea
for $r>r_b$ (here $\tilde v_b\equiv |\tilde{\bf v}_b|$). The absolute value of the resulting ratio
\ba{21} &{}&\frac{\Phi_{vb}}{\tilde\Phi_b}=\frac{3\mathcal{H}\lambda \tilde{v}_b}{a}\frac{\rho_b}{\rho_b-\overline{\rho}}\nn\\
&\times&\frac{\lambda}{ar}\left\{1+\frac{ar}{\lambda}-\frac{1}{3}\left(\frac{ar_b}{\lambda}\right)^3\left[\frac{ar_b}{\lambda}\cosh\left(\frac{ar_b}{\lambda}\right)-\sinh\left(\frac{ar_b}{\lambda}\right)\right]^{-1}\exp\left(\frac{ar}{\lambda}\right)\right\}\nn\\
&{}&\ea
is now not only directly proportional to the small factor $3\mathcal{H}\lambda \tilde{v}_b/a$, but also inversely proportional to the factor $(\rho_b-\overline{\rho})/\rho_b$, which decreases when the size of the ball increases (as a manifestation of the cosmological principle, $\rho_b\approx\overline{\rho}$ when the scale $r_b$ is large enough while still loosely fitting in the simulation box). Hence, owing to mutual compensation of these two multipliers, obliteration of the velocity-dependent contribution $\Phi_{vb}$ from the total scalar perturbation $\Phi_b=\tilde\Phi_b+\Phi_{vb}$ at sufficiently large scales is placed under taboo.

\

\section{Screening length in linear perturbation theory}
\label{Sec3}

Whereas the previous section has been devoted to the screening of gravity in the inhomogeneous Universe, predicted by discrete cosmology, here we are going to concentrate on the radically different screening mechanism in the framework of the relativistic perturbation theory applicable in the large-scale spatial regions where the energy density fluctuation $\delta{\varepsilon}(\eta,{\bf r})\equiv\varepsilon-\overline{\varepsilon}$ is small as compared to $\overline{\varepsilon}$. Disregarding vector and tensor perturbations, we start with the corresponding linearized Einstein equations \cite{Rubakov,Mukhanov,Bernardeau}:
\be{22}\triangle\Phi-3\mathcal{H}\left(\Phi'+\mathcal{H}\Phi\right)=\frac{1}{2}\kappa a^2\delta\varepsilon\, ,\ee
\be{23}\Phi'+\mathcal{H}\Phi=-\frac{1}{2}\kappa a^2\overline\varepsilon \nu\, , \ee
\be{24}\Phi''+3\mathcal{H}\Phi'+\left(2\mathcal{H}'+\mathcal{H}^2\right)\Phi=0\, ,\ee
where $\nu(\eta,{\bf r})$ is the velocity potential. Following \cite{Hahn}, we assume that
\be{25}\Phi=\frac{D_1}{a}\phi\, ,\ee  
where the introduced function $\phi(\mathbf{r})$ does not depend on $\eta$, and $D_1(\eta)$ denotes the so-called linear growth factor. Substitution of \rf{25} into Eqs.~\rf{23} and \rf{24} gives
\be{26} D_1'\phi=-\frac{1}{2}\kappa a^3 \overline\varepsilon \nu\, ,\ee
\be{27} D_1''+\mathcal{H}D_1'+\left(\mathcal{H}'-\mathcal{H}^2\right)D_1=0\, ,\ee
respectively. The latter equation admits two independent solutions (given by Eqs.~(29) and (31) in \cite{Bernardeau}):
\be{28}D_1^{(+)}\propto H\int\frac{da}{\left(aH\right)^3} \propto \frac{\mathcal{H}}{a}\int\frac{da}{\mathcal{H}^3}\, , \ee
\be{29}D_1^{(-)}\propto \frac{\mathcal{H}}{a}\, .\ee

As a consequence of \rf{25},
\be{30}
\Phi'+\mathcal{H}\Phi=\frac{D_1'}{a}\phi=\frac{D_1'}{D_1}\Phi\, .
\ee
Substituting \rf{30} into Eq.~\rf{22}, we derive Eq.~(16) from \cite{Hahn} (up to notation),
\be{31}\triangle\Phi-3\mathcal{H}\frac{D_1'}{D_1}\Phi=\frac{1}{2}\kappa a^2\delta\varepsilon\, , \ee
along with the corresponding comoving screening length given by Eq.~(17) from \cite{Hahn},
\be{32} l\equiv\frac{1}{\sqrt{3\mathcal{H}^2f}}\, ,\quad f\equiv\frac{d\ln D_1}{d\ln a}\, .\ee

Of course, it is now absolutely necessary to briefly contrast the Helmholtz equations \rf{3} and \rf{31}. In \rf{3} there are two sources of $\Phi$: one $\propto\delta\rho$ (producing the compact velocity-independent part $\tilde\Phi$ \rf{13}) and the other $\propto\Xi$ (generating the cumbersome velocity-dependent part $\Phi_v$ \rf{17}). In \rf{31} there is only one source $\propto\delta\varepsilon$. Hence, in Eq.~\rf{31} the contribution of $\Phi$ to $\delta\varepsilon$ is left out of account. At the same time, figuratively speaking, the velocity potential $\nu$ is converted into the scalar perturbation $\Phi$ itself (see Eqs.~\rf{23}, \rf{30}), ensuring the screening of gravity with the characteristic comoving range $l(\eta)$ \rf{32}. On the contrary, in Eq.~\rf{3} the inconvenient velocity-dependent source is left unconverted, but $\Phi$ is singled out from $\delta\varepsilon$ \cite{Eingorn,BrilEinPDU}:
\be{33} \delta\varepsilon=\frac{c^2}{a^3}\delta\rho+\frac{3\overline\rho c^2}{a^3}\Phi\, .\ee

For the considered large-scale spatial regions $\Xi=\overline{\rho}\nu$, and Eq.~\rf{3} takes the form
\ba{34} \triangle\Phi - \frac{3\kappa\overline\rho c^2}{2a} \Phi = \frac{\kappa c^2}{2a}\delta\rho-\frac{3\kappa \overline\rho c^{2}{\mathcal H} }{2a} \nu\, . \ea
Substitution of \rf{25} and \rf{26} into Eq.~\rf{34} gives
\be{35}\delta\rho=\frac{2D_1}{\kappa c^2}\triangle\phi-\left(\frac{6\mathcal{H} D_1'}{\kappa c^2} + \frac{3\overline\rho D_1}{a}\right) \phi\, ,\ee
with the corresponding Fourier transform
\be{36}
\hat{\delta\rho}=-\frac{2}{\kappa c^2}\left(D_1k^2+3\mathcal{H} D_1'+\frac{3\kappa \overline\rho c^2D_1}{2a}\right)\hat{\phi}\, .
\ee

The time has come to keep our promise and demonstrate the great importance of the term $\propto\nu$ in Eq.~\rf{34} for the correct description of structure growth (in addition, see \cite{Dent}). For the sake of simplicity, we confine ourselves to the matter-dominated evolution stage when $\mathcal{H}^2=\kappa\overline{\rho}c^2/(3a)$. Concentrating exclusively on the growing mode $D_1^{(+)}\propto a$ (see Eq.~\rf{28}), from \rf{36} we get
\be{37} \hat{\delta\rho}\propto k^2\hat{\phi}\left(a+\frac{5\kappa\overline\rho c^2}{2k^2}\right)\, .\ee
This function represents the dominant solution of Eq.~(30) from \cite{BrilEinPDU}, derived specifically for $\hat{\delta\rho}(\eta,k)$ with the regard for the cosmological screening. On the contrary, if one obliterated the term $\propto\nu$ in Eq.~\rf{34}, instead of \rf{37} we would get
\be{38} \hat{\widetilde{\delta\rho}}\propto k^2\hat{\phi}\left(a+\frac{3\kappa\overline\rho c^2}{2k^2}\right)\, .\ee
With all due respect, this function does not satisfy Eq.~(30) from \cite{BrilEinPDU} and is unsuitable unless $a\gg\kappa\overline{\rho} c^2/k^2$ (this inequality holds true at sufficiently small distances where the cosmological screening does not come into play, so the expressions \rf{37} and \rf{38} coincide with each other).

\

\section{Effective screening length to rule them all}
\label{Sec4}

Evidently, the temptation to combine the analyzed screening mechanisms is too strong. The source $\propto\Xi$ in Eq.~\rf{3} is insignificant at small enough scales \cite{Eingorn}. Consequently, it may be safely replaced by the term $\propto\nu$, resulting in Eq.~\rf{34}. Such a trick is well-grounded, since the non-linearity scale ($\sim15\,\mathrm{Mpc}$ today \cite{Villa}), where the linear perturbation theory fails, is much less than the investigated screening ranges. Thus, at those very distances, where $\Xi$ may not be accurately approximated as $\overline{\rho}\nu$, this nuisance is of no importance, because in the small-scale spatial regions the whole source $\propto\Xi$ may be totally ignored along with the term $\propto\Phi$, leaving us alone with the standard Poisson equation $\triangle\Phi = \left(\kappa c^2/2a\right)\delta\rho$ of Newtonian cosmology. Returning to Eq.~\rf{34} and expressing $\nu$ via $\Phi$ with the help of Eqs.~\rf{23}, \rf{30}, we immediately derive the Helmholtz equation,  deemed appropriate at arbitrary distances:
\be{39}\triangle\Phi-\frac{a^2}{\lambda_{\mathrm{eff}}^2}\Phi=\frac{\kappa c^2}{2a} \delta\rho\, ,
\ee 
where $\lambda_{\mathrm{eff}}(\eta)$ is the effective physical (i.e. non-comoving) screening length introduced via the equality
\be{40} \frac{1}{\lambda_{\mathrm{eff}}^2}\equiv \frac{1}{\lambda^2}+\frac{1}{a^2l^2}\, .\ee
The same result follows from Eq.~\rf{31} after substitution of the formula \rf{33}. The Helmholtz equation \rf{39} inherits the advantages of both its predecessors \rf{3} and \rf{31}: there is only one velocity-independent source $\propto\delta\rho$, analytically determined by the positions of gravitating masses (see Eq.~\rf{5}).

\begin{figure}[!ht]
	\centering
	\includegraphics{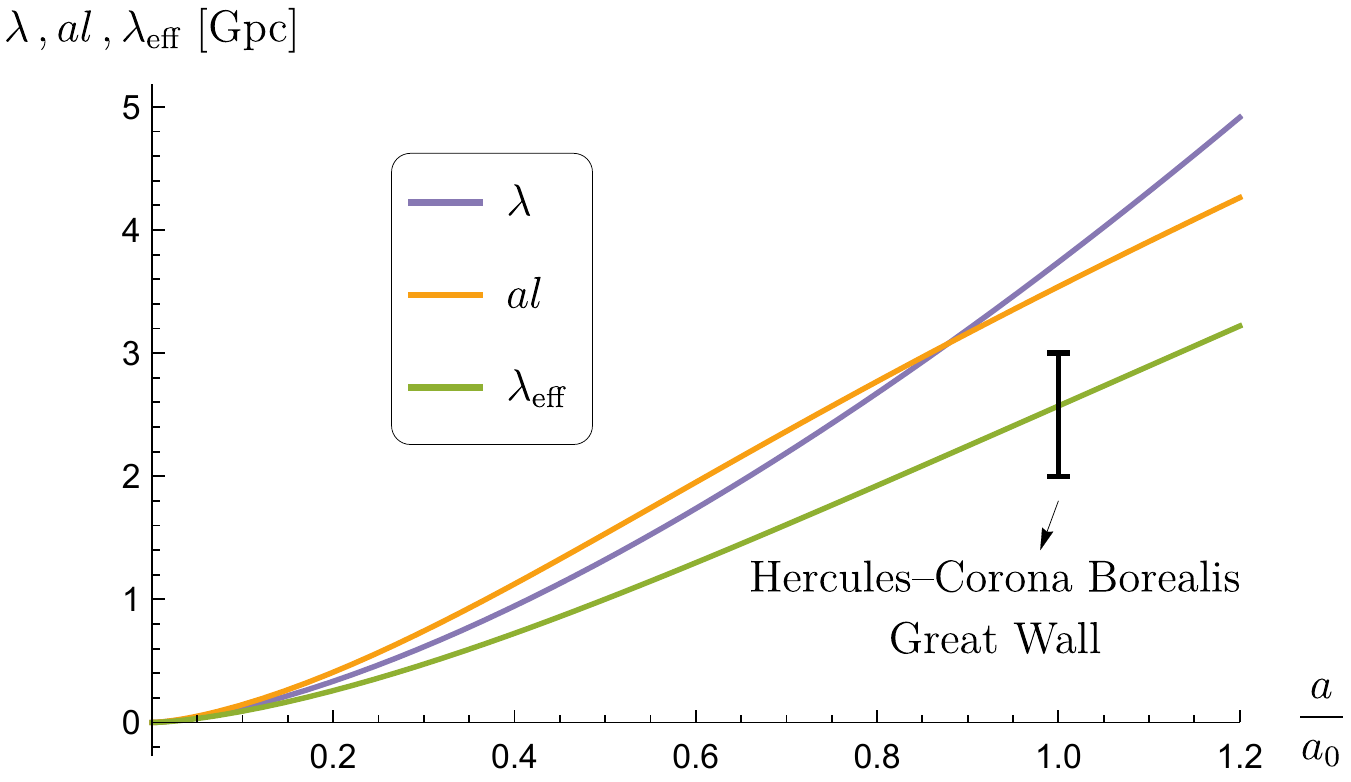}
	\caption{Screening lengths $\lambda$, $al$ and $\lambda_{\mathrm{eff}}$ as functions of the normalized scale factor $a/a_0$.}
	\label{Figure1}
\end{figure}

Focusing again on the growing mode $D_1^{(+)}$ \rf{28} and using the definitions \rf{9} and \rf{32}, as well as the Friedmann equations \rf{2}, we find
\be{41}
\frac{1}{\lambda_{\mathrm{eff}}^2}=\frac{3}{a\mathcal{H}}\left(\int \frac{da}{\mathcal{H}^3}\right)^{-1}=\frac{3}{c^2a^2H}\left(\int \frac{da}{a^3H^3}\right)^{-1}\, ,
\ee
where the Hubble parameter
\be{42} H=H_0\sqrt{\Omega_\mathrm{M}\left(\frac{a_0}{a}\right)^3+\Omega_{\Lambda}}\, , \quad \Omega_\mathrm{M}\equiv \frac{\kappa\overline\rho c^4}{3H_0^2a_0^3}\, , \quad \Omega_{\Lambda}\equiv \frac{\Lambda c^2}{3H_0^2}\, .\ee
Here $a_0$ is the current value of the scale factor. Finally,
\be{43} \lambda_{\mathrm{eff}}=\sqrt{\frac{c^2a^2H}{3}\int \frac{da}{a^3H^3}}\, .\ee
Relying on the cosmological parameters $H_0= 67.4\,\mathrm{km}\,\mathrm{s}^{-1}\,\mathrm{Mpc}^{-1}$, $\Omega_\mathrm{M}=0.315$, $\Omega_{\Lambda}=0.685$ \cite{Planck}, we depict the temporal dependence of $\lambda$ \rf{9}, $al$ (see \rf{32}) and $\lambda_{\mathrm{eff}}$ \rf{43} in Fig.~\ref{Figure1}. Today $\lambda_0=3.74\,\mathrm{Gpc}$, $(al)_0=3.54\,\mathrm{Gpc}$ and $\left(\lambda_{\mathrm{eff}}\right)_0=2.57\,\mathrm{Gpc}$.

It is interesting to note that during the matter-dominated stage of the Universe evolution, when $H^2=H_0^2\Omega_{\mathrm{M}}\left(a_0/a\right)^3$ (and, therefore, $1/H\propto a^{3/2}$), the following equalities and inequalities hold true:
\be{44}
\lambda_{\mathrm{eff}}=\sqrt{\frac{2}{15}}\frac{c}{H}\, ,\quad \lambda=\frac{\sqrt{2}}{3}\frac{c}{H}\, ,\quad al=\frac{1}{\sqrt{3}}\frac{c}{H}\, ,\quad \lambda_{\mathrm{eff}}<\lambda<al\, .
\ee

According to \rf{37}, at this evolution stage $\hat{\delta\rho}$ substantially grows below the characteristic comoving scale $k^{-1}=\sqrt{2a/(5\kappa\overline{\rho}c^2)}=\lambda_{\mathrm{eff}}/a$. This means that, strictly speaking, $\lambda_{\mathrm{eff}}$ (and not $\lambda$, as claimed in \cite{Eingorn,BrilEinPDU}, although the values of $\lambda_{\mathrm{eff}}$ and $\lambda$ are of the same order) defines the size of a spatial domain where cosmic structures may grow. A hypothesis offered in \cite{Eingorn} interprets $\lambda$ as the upper bound for the dimensions of a solitary structure. Now we reinforce this hypothesis by assigning this role to $\lambda_{\mathrm{eff}}$ instead. The made assignment fully agrees with the observational data: the effective cosmological screening length $\left(\lambda_{\mathrm{eff}}\right)_0\approx2.6\,\mathrm{Gpc}$ exceeds the diameter of Giant GRB Ring $\sim1.7\,\mathrm{Gpc}$ \cite{ring,Balazs} and perfectly matches the size of Hercules--Corona Borealis Great Wall (Her--CrB GW) $\sim2\div3\,\mathrm{Gpc}$ \cite{wall,wall2} (see Fig.~\ref{Figure1}). Provided that Her--CrB GW of dimension $\sim\left(\lambda_{\mathrm{eff}}\right)_0$ does exist and the advanced hypothesis is true, this colossal structure may be called not just ``the largest observed'', but simply ``the largest'' in the Universe.

Finally, let us present the exact solution of Eq.~\rf{39} for discrete gravitating masses:
\be{45}
\Phi=\frac{1}{3}\left(\frac{\lambda_{\mathrm{eff}}}{\lambda}\right)^2 -\frac{\kappa c^2}{8\pi a}\sum_n\frac{m_n}{|\mathbf{r}-\mathbf{r}_n|}\exp\left(-\frac{a|\mathbf{r}-\mathbf{r}_n|}{\lambda_{\mathrm{eff}}}\right)\, .
\ee
This analytical expression inherits the advantages of the scalar perturbation $\Phi$ \rf{7} investigated in \cite{Eingorn}. First, it diverges only at positions of particles and nowhere else. Second, its average value is equal to zero (see Eqs.~(3.13) and (3.14) in \cite{Eingorn}):
\ba{46} \overline\Phi\equiv \frac{1}{\mathcal{V}}\int\limits_{\mathcal{V}}d{\bf r}\Phi&=&\frac{1}{3}\left(\frac{\lambda_{\mathrm{eff}}}{\lambda}\right)^2 -\frac{\kappa c^2}{8\pi a}\frac{1}{\mathcal{V}}\int\limits_{\mathcal{V}}d{\bf r}\sum_n\frac{m_n}{|\mathbf{r}-\mathbf{r}_n|}\exp\left(-\frac{a|\mathbf{r}-\mathbf{r}_n|}{\lambda_{\mathrm{eff}}}\right)\nn\\
&=&\frac{1}{3}\left(\frac{\lambda_{\mathrm{eff}}}{\lambda}\right)^2 -\frac{\kappa c^2}{8\pi a}\,\overline\rho\,\frac{4\pi \lambda_{\mathrm{eff}}^2}{a^2}=0\, ,\ea
where $\mathcal{V}$ stands for the infinite comoving averaging volume, $(1/\mathcal{V})\sum\limits_nm_n\equiv\overline\rho$.

Last but not least, the formula \rf{45} is valid for arbitrary distances and ensures Newtonian gravitational interaction at sub-horizon scales as well as in the Minkowski background limit. This formula is much more compact and convenient than \rf{7} for the cosmological simulation purposes. It is ready to be used in the equation of motion \rf{12} instead of $\tilde\Phi$, without fear of unreliable predictions in the domain of linear density fluctuations, and at the same time with confidence in the description below the non-linearity scale.

\ 

\section{Conclusion}
\label{Sec5}

In this paper we have organized a bloodless duel of the cosmological screening lengths: $\lambda$ \rf{9}, springing from the corresponding Helmholtz equation \rf{3} studied in \cite{Eingorn}, and $al$ (see \rf{32}), springing from the other Helmholtz equation \rf{31} investigated in \cite{Hahn}. We must frankly confess that initially we seconded the formalism of discrete cosmology \cite{Eingorn}. However, very soon we realized that the desirable neglect of the source $\propto\Xi$ in Eq.~\rf{3}, that would substantially simplify the solution (one can easily compare the explicit expressions \rf{7} and \rf{13}), would simultaneously make the developed theory unsuitable at large enough scales (\rf{37} versus \rf{38}). Therefore, we have willingly resorted to the formalism of \cite{Hahn} and replaced the velocity-dependent term $\propto\Xi$ by the term $\propto\Phi$, thereby giving rise to the novel Helmholtz equation \rf{39}. It determines the scalar perturbation $\Phi$ at all cosmic scales including the domain of nonlinear density contrasts (see the exact analytical solution \rf{45}, which is much simpler than \rf{7} and can be widely used in cosmological simulations).

Thus, as it sometimes happens, the real winner of the confrontation is the third party, namely, the effective screening length $\lambda_{\mathrm{eff}}$ \rf{43}. The structure formation is suppressed at distances above this finite time-dependent Yukawa range. In particular, its current value $\left(\lambda_{\mathrm{eff}}\right)_0\approx2.6\,\mathrm{Gpc}$ coincides with the size of Her--CrB GW, being the most gigantic known object in our bewitching Universe.

\ 


\end{document}